# First-Principles Study Lead-Free Halide Double Perovskite $Cs_2RhAgX_6$ and $Cs_2IrAgX_6$ (X = Cl, Br and I)

Yue Kuai[a], Changcheng Chen[a,*], Pengfei Lu[b,*]

[a] *School of Science, Xi'an University of Architecture and Technology, Xi'an 710055, Shaanxi, China*

[b] *State Key Laboratory of Information Photonics and Optical Communications, Beijing University of Posts and Telecommunications, Beijing 100876, China*

[c] *Laboratory of Plasma and Energy Conversion, School of Physics and Optoelectronic Engineering, Ludong University, Yantai 264025, China*

**ABSTRACT**

In contrast to lead-based perovskites, double perovskites have attracted considerable interest due to their ability to modulate photovoltaic properties and high stability through elemental control. However, most double perovskites are mainly faced with large band gap ranges and indirect band gaps. Here, we report the structural, electronic, and optical properties of the double perovskites $Cs_2RhAgX_6$ and $Cs_2IrAgX_6$ (X = Cl, Br and I) by first-principles calculations. The results of the thermodynamic stability and electronic properties show that the double perovskites have a higher stability and exhibit a suitable band gap range in the field of optoelectronics, their band gap decrease with the substitution of halogen elements from Cl to I (0.55-2.2 eV). In addition, the double perovskite exhibits strong light absorption. These results suggest a great potential for $Cs_2RhAgX_6$ and $Cs_2IrAgX_6$ (X = Cl, Br and I) perovskites in optoelectronic applications.

1. INTRODUCTION

Lead-based halogen perovskite APbX$_3$ (A = CH$_3$NH$_3$, (H$_2$N)$_2$CH, Cs, X = Cl, Br and I) has been widely investigated in photovoltaic absorbers, light-emitting diodes, and photodetectors due to its excellent optoelectronic properties and low fabrication cost [1-3]. At present, the most investigated organic-inorganic halide perovskite MAPbI$_4$ has achieved the highest photoelectric conversion efficiency of 25.4% in solar devices [4]. However, there are two main challenges for large-scale commercial application of lead-based perovskite. The first is that lead harms human health and the environment. The second is that mixed perovskites tend to degrade in humidity, oxygen, and high temperatures. These challenges promote the search for highly stable lead-free halogen perovskite materials [5,6].

In an effort to address the toxicity and instability of Pb, efforts have been made to replace Pb with the Sn and Ge. Since the divalent states of Ge and Sn have a similar electronic configuration to Pb, this would maintain the high electronic dimension and three-dimensional perovskite structure consistent with the Pb-based halides perovskites [7]. The divalent Ge and Sn based halide perovskites have more serious stability issues as they are easily oxidized to the tetravalent state [8]. In addition, Sn may be more harmful to the environment and living organisms than Pb. Within this context, an effective solution is to replace the lead with monovalent and trivalent cations to form the halogen double perovskite Cs$_2$BB'X$_6$. The double perovskite has the same octahedral connectivity as ABX$_3$, with [B$^+$X$_6$] and [B'$^{3+}$X$_6$] octahedra alternating throughout space, so that the connectivity of the same octahedron is essentially zero-dimensional (0D) [9]. The combination of univalent cations and trivalent cations makes double perovskite have good heat and moisture resistance due to the fact that most of the elements that make up double perovskite can maintain stable univalent and trivalent oxidation states [10]. Due to the abundance of B$^+$ and B'$^{3+}$ elements, the diversity of double perovskites materials is increased and they also exhibit high tunability in terms of optical and electronic properties [11].

However, the double perovskite energy band structure is mainly determined by the electronic orbital symmetry of the B- and B'-site elements.[12] The large indirect band gaps and electron-hole masses exhibited by most double perovskite due to the asymmetry of the B- and B'-site electron orbitals limit the further development of double perovskite in the field of optoelectronic devices. Although combinations of Na$^+$ and In$^{3+}$, Ag$^+$ and Sb$^{3+}$, In$^+$ and Sb$^{3+}$ ions have been reported to exhibit electronic properties similar to those of Pb-based perovskite, the large direct band gap created by the combination of $Na^+$ and $In^{3+}$ is not suitable for use as a solar cell [12-17]. In addition, the $In^+$ is not easily oxidized to $In^{3+}$, and Sb/Tl is more harmful to humans and the environment than Pb. On this basis, we need to design a stable halogen double perovskite with a band gap suitable for optoelectronic devices. For most double perovskites, $B^{3+}$ is usually an element of IIIA (In and Tl) and VA (Sb and Bi), while the main-group elements further up the periodic table (Rh and Ir) have been less studied [18-20]. In addition, X-site halogen substitution enhances double perovskites stability and band gap, and optical properties [21-25].

In this work, we have designed $Cs_2B'AgX_6$ (B' = Rh, Ir, X = Cl, Br and I) halide double perovskites by performing elemental substitutions in the same main group at the B' and X positions, and have explored the effects of elemental substitutions at the B' and X positions in terms of structural stability, electronic and optical properties, and the range of applications for which they are suitable. Firstly, the changes in structure, perovskite phase stability, thermodynamic stability and mechanical properties due to elemental substitution of B' and X were investigated by calculating bond lengths, octahedral and Goldsmith tolerance factors, enthalpies of formation and decomposition and Young's modulus for B'-X and Ag-X. More, the electronic properties of $Cs_2RhAgX_6$ and $Cs_2IrAgX_6$ (X = Cl, Br and I), including energy bands, density of states and charge projections, are systematically investigated using the HSE06 and HSE06+SOC functions. Finally, the dielectric constants of $Cs_2RhAgX_6$ and $Cs_2IrAgX_6$ (X = Cl, Br and I) have been calculated to explore the effect of elemental substitution at the B' and X positions on the optical properties and thus determine their scope of application in optoelectronic devices. Our work is expected to lead to important potential applications of $Cs_2RhAgX_6$ and $Cs_2IrAgX_6$ (X = Cl, Br and I) in optoelectronic devices.

## 2. COMPUTATIONAL DETAILS

All calculations are performed using Density Functional Theory (DFT) as implemented in the PWmat simulation package [26,27]. The exchange-correlation function is described by using the generalized gradient approximation (GGA) of the Perdew-Burke-Ernzerhof (PBE) formula with SG15 norm-conserving pseudopotential. The cut-off energy of the plane wave was set to 350 eV [28,29]. The 7 × 7 × 7 k-point mesh for Brillouin zone integration is employed for the unit cells. For lattice relaxation, the convergence criterion for force and the convergence criterion for energy are 0.01 eV

Å$^{-1}$ and 10$^{-5}$ eV [30]. Characterization of electron-ion interactions using the all-electron projector-augmented wave (PAW) method [31]. The energy band structure, density of states and optical properties of the system were calculated using the hybrid functional Heyd-Scuseria-Ernzerhot (HSE06) due to the fact that PBE functions usually underestimate the semiconductor and insulator band gaps [32,33]. When calculating the electronic properties of the system, the spin-orbit coupling (SOC) effect is considered due to the presence of heavy elements in the system. High symmetry k-path W (0.50,0.25,0.75), L (0.50,0.50,0.50), Γ (0.0,0.0,0.0), X (0.50,0.00,0.50), W (0.50,0.25,0.75) were used for band structure calculations.

## 3. RESULTS AND DISCUSSION

### 3.1 Geometry and stability

The **Fig. 1a** shows the AMX$_3$ halogen perovskite structure consisting of a common-angled halide [M$^{2+}$X$_6$] octahedron. The A site is occupied by univalent cations (e.g Cs$^+$), which occupies a cubic octahedral cavity composed of eight common-angle metal halide octahedrons. The M site is usually divalent metal cations (e.g Pb$^{2+}$), and the X site is halogen anions (e.g Cl$^-$, Br$^-$ and I$^-$). For the double perovskite structure of A$_2$B'BX$_6$, the M site consists of horizontally and vertically alternating B$^+$ ions and B'$^{3+}$ ions, as shown in **Fig. 1b**. The **Fig. 1c** shows the rhomboid unit cell structure of double perovskite, which is different from the standard cubic cell axis at 90°. The angle between the rhomboid cell axes is 60°. All structures maintain face-centered cubic symmetry (space group Fm3m) [12]. We optimized the structures of six candidate materials Cs$_2$B'AgX$_6$ (B' = Rh, Ir, X = Cl, Br and I) for B'$^{3+}$ and X$^-$ ion combinations using the PBE functional with the lattice parameters shown in **Fig. 1d-f**. The effect of the X-site halogen element on the lattice constant, Ag-X bond length and B'-X bond length is significantly different from that of the B'-site element, which has almost negligible effect on the lattice constant and bond length. The lattice constants of Cs$_2$B'AgX$_6$ (B' = Rh, Ir, X = Cl, Br and I), Ag-X bond length and B'-X bond length increase with the substitution of halogen elements from Cl to I due to the electronegativity of the halogen ion: I$^-$<Br$^-$<Cl$^-$.

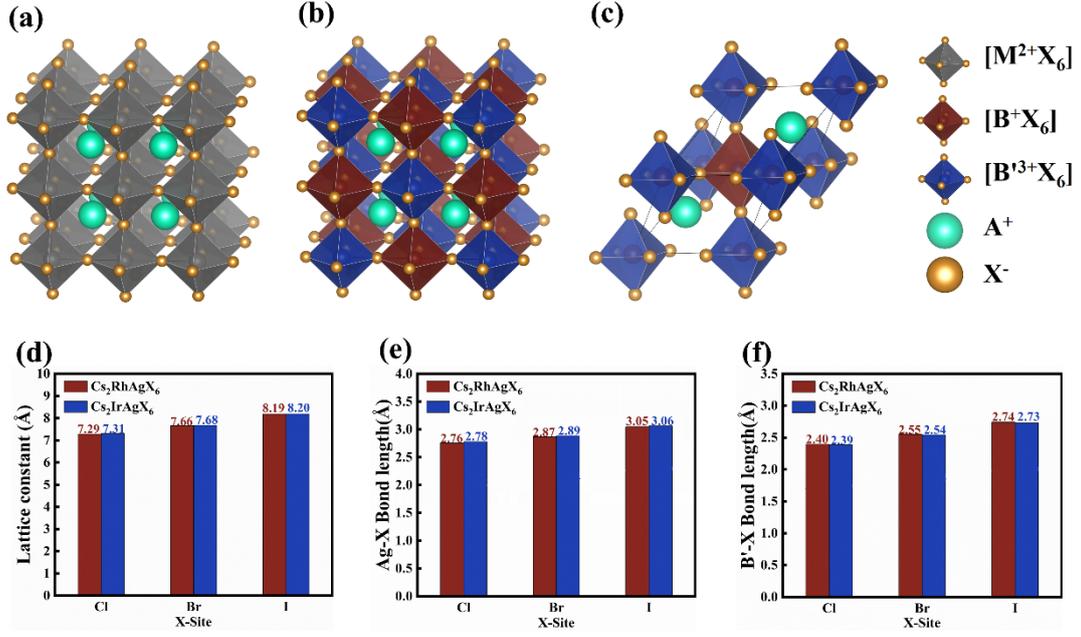

**Fig. 1** (a) Typical inorganic halogenated AMX₃ perovskite standard structure, (b) inorganic halogenated A₂B'BX₆ perovskite standard structure and (c) inorganic halogenated A₂B'BX₆ perovskite primary cell structure. The variation of (d) lattice constants, (e) Ag-X bond lengths, (f) B'-X bond lengths in relation to the substitution of elements at the X and B' positions (B' = Rh, Ir, X = Cl, Br and I).

In order to further explore the stability of $Cs_2B'AgX_6$ perovskite (B' = Rh, Ir, X = Cl, Br and I), we considered the tolerance factor $t$ and octahedral factor $\mu$ to analyze the geometry. For the double perovskite structure of $Cs_2B'AgX_6$, the tolerance factor $t$ is defined as follows [34]:

$$t = \frac{r_{Cs} + r_X}{\sqrt{2}(r_M + r_X)} \tag{1}$$

The $r_{Cs}$ and $r_X$ represent the Shannon ion radius of A-site cation and X-site halogen ion. The $r_M = [(r_{B'} + r_{Ag})/2]$, which $r_{B'}$ and $r_{Ag}$ are Shannon ion radius of B'-site (Rh, Ir) and Ag ion.

For ideal perovskite structures, the tolerance factor $t$ has a value of 1. Highly stable cubic phase structures are maintained when the tolerance factor $t$ is in the range $0.81 < t < 1.1$. When the value of the tolerance factor $t$ is small, the cooperative octahedral deformation of halide peroxides can lead to tetragonal, orthogonal phases, or even hexagonal shapes with low symmetry. On the other hand, larger tolerance factor $t$ values may lead to a distortion of the perovskite structure, from a three-dimensional crystal to a layered structure. For the double perovskite structure, the octahedral factor μ is used to determine B'X$_6$ and AgX$_6$ octahedra stability in Cs$_2$B'AgX$_6$ perovskite (B' = Rh, Ir, X = Cl, Br and I), and the formula is [34]:

$$\mu = \frac{r_M}{r_X} \quad (2)$$

For halide perovskite, the judgment standard is 0.44 < μ < 0.9. Experimental results show that unstable perovskite with octahedral factor μ less than 0.442 can be synthesized even if the structure has a reasonable octahedral factor μ [14,33]. In **Fig. 2a**, we summarized the tolerance factor $t$ and octahedral factor μ for Cs$_2$B'AgX$_6$ perovskite (B' = Rh, Ir, X = Cl, Br and I) to explore the stability of its cubic phase structure in relation to B' ions and halogen X ions. The B' site ion substitution has less effect on the tolerance factor $t$ and octahedral factor μ than the halogen ion substitution. Replacement of Cl⁻ ions effectively improves cubic phase stability compared to Br⁻ and I⁻ ions.

In order to further evaluate the thermodynamic stability of Cs$_2$B'AgX$_6$, we explored their formation energy $E_f$ and decomposition enthalpy $\Delta H_d$ [35,36].

$$E_f = \frac{[E_{Cs_2MAgX_6} - (2E_{Cs} + E_{Ag} + E_M + 6E_X)]}{10} \quad (3)$$

We further investigated the stability of peroxides by the PBE function, and we calculated the decomposition enthalpy $\Delta H_d$ along $Cs_2B'AgX_6 \rightarrow B'X_3 + 2CsX + AgX$ decomposition path [35,36].

$$\Delta H_d = \frac{[E_{Cs_2MAgX_6} - (2E_{CsX} + E_{AgX} + 6E_{B'X_3})]}{10} \quad (4)$$

where $E_{Cs}$, $E_{Ag}$, $E_{B'}$, and $E_X$ are the average energy of each atom in pure solid simple substance for Cs, B', Ag and X respectively, after structural optimization and $E_{Cs_2B'AgX_6}$ represent the energy of Cs$_2$B'AgX$_6$ perovskite (B' = Rh, Ir, X = Cl, Br and I). The $E_{CsX}$, $E_{AgX}$ and $E_{B'X_3}$ represent CsX, AgX and B'X$_3$ unit cell energy respectively.

The formation energy $E_f$ and decomposition enthalpy $\Delta H_d$ of $Cs_2AgRhX_6$ and $Cs_2AgIrX_6$ (X = Cl, Br and I) perovskite are shown in **Fig. 2b** and **Fig. 2c**, respectively. The negative $E_f$ and $\Delta H_d$ mean that the crystals are thermodynamically stable, the crystal formation process is exothermic and the decomposition process is endothermic [37-39]. In addition, the bond energies of the Cs-Cl, Ag-Cl and B'-Cl bonds are higher than those of the Cs-Br/I, Ag-Br/I and B'-Br/I bonds. Therefore, $Cs_2AgB'Cl_6$ is more stable than $Cs_2AgB'Br_6$ and $Cs_2AgB'I_6$ for the X-position halogen ion substitution (B' = Rh, Ir). For the B'-site substitution, $Cs_2AgRhX_6$ is slightly more stable than $Cs_2AgIrX_6$ (X = Cl, Br and I). The Young's modulus of the three materials is also shown in **Fig. 2d**, the electronegativity of the halogen element also affects the mechanical properties of the $Cs_2 B'AgX_3$ perovskite (B' = Rh, Ir, X = Cl, Br and I).

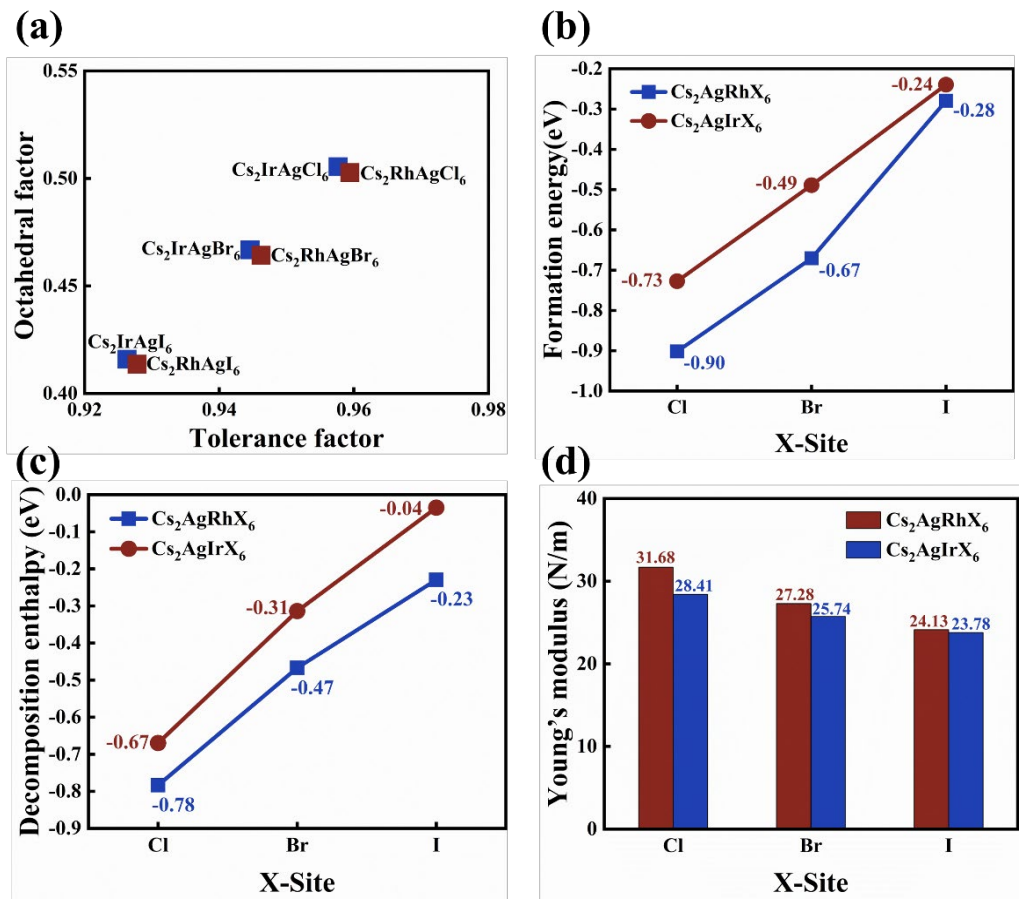

**Fig. 2** (a) Tolerance factor $t$ and octahedral factor μ, (b) formation energy $E_f$, (c) decomposition enthalpy $\Delta H_d$ and (d) Young's modulus of $Cs_2B'AgX_6$ perovskite (B' = Rh, Ir, X = Cl, Br and I).

## 3.2 Electronic properties

To determine the suitable range of applications for $Cs_2RhAgX_6$ and $Cs_2IrAgX_6$ (X = Cl, Br and I), their band structures and density of states (DOS) are calculated. As shown in **Fig. 3a** and **Fig. 3b**, the band gaps of $Cs_2RhAgX_6$ and $Cs_2IrAgX_6$ (X = Cl, Br and I) are calculated by using HSE06 and HSE06+SOC functional. The band gaps decrease monotonically as substitution of halogen elements from Cl to I, this is attributed to more electronegative X-site anions which leads to a lower-energy valence band (**Fig. 3a**). The band gap range are 0.77-2.12 eV for $Cs_2RhAgX_6$ and $Cs_2IrAgX_6$ (X = Cl, Br and I). Considering the effects of spin-orbit coupling (SOC) due to HSE06 functional approximation, which carries uncertainties in predicting the band gap [40-42]. In **Fig. 3b,** the HSE06+SOC functional is applied to calculate the band gaps of $Cs_2RhAgX_6$ and $Cs_2IrAgX_6$ (X = Cl, Br and I). Results show that the HSE06+SOC functional calculation is 0.55-1.88 eV, which is lower than that of HSE06 functional calculation. According to the Shockley-Queisser limit, the solar cell assembled by a semiconductor with a band gap of 1.0-1.6 eV obtains the maximum efficiency of 30-33% [43-46]. The band gaps of $Cs_2RhAgBr_6$, $Cs_2IrAgCl_6$, $Cs_2IrAgBr_6$ and $Cs_2IrAgI_6$ are 0.88-1.6 eV (**Fig.3b**), which are suitable for solar cell application.

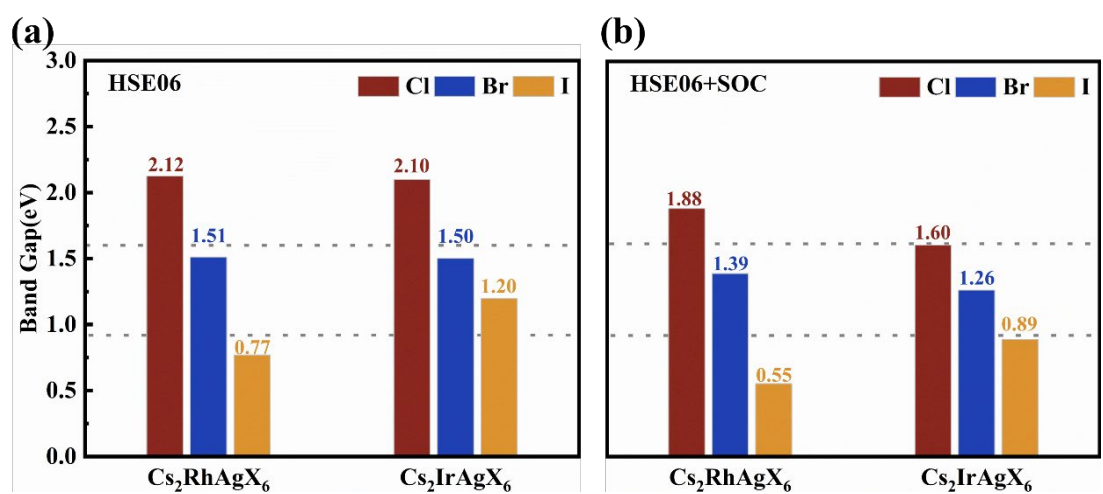

**Fig. 3** The band gaps of $Cs_2RhAgX_6$ and $Cs_2IrAgX_6$ (X = Cl, Br and I) are calculated by using (a) HSE06 and (b) HSE06+SOC functional. The red, blue and orange distributions are corresponded to Cl, Br and I in $Cs_2RhAgX_6$ and $Cs_2IrAgX_6$ systems at the X-site.

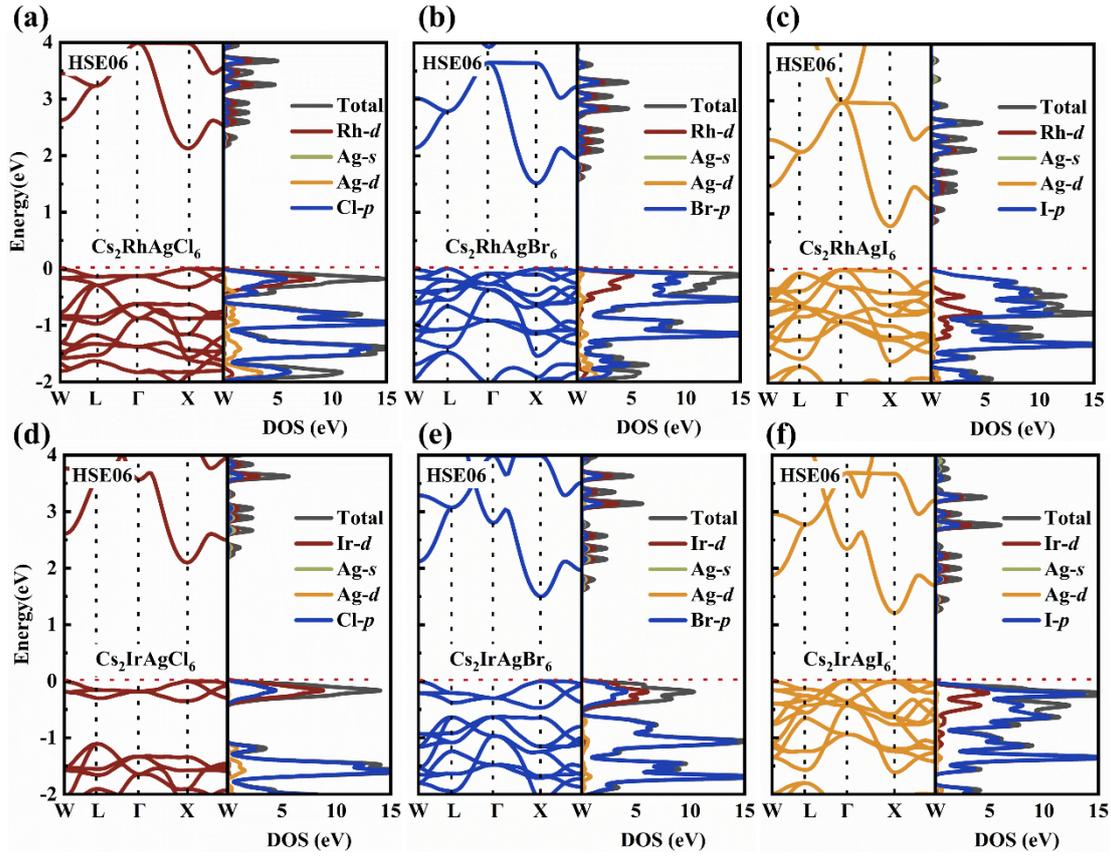

**Fig. 4** HSE06 functional calculated band structures and TDOS/PDOS for the $Cs_2B'AgX_6$ (B' = Rh, Ir, X = Cl, Br and I) perovskites: (a) $Cs_2RhAgCl_6$, (b) $Cs_2RhAgBr_6$, (c) $Cs_2RhAgI_6$, (d) $Cs_2IrAgCl_6$, (e) $Cs_2IrAgBr_6$ and (f) $Cs_2IrAgI_6$. The fermi-level is set at zero.

**Fig. 4a-f** display the band structures and total/partial density of states (TDOS/PDOS) for the $Cs_2B'AgX_6$ (B' = Rh, Ir, X = Cl, Br and I) perovskites, which are calculated by using HSE06 functional. The high symmetry point coordinates are marked as W (0.50, 0.25, 0.75), L (0.50, 0.50, 0.50), Γ (0.0, 0.0, 0.0) and X (0.50, 0.00, 0.50). The $Cs_2RhAgCl_6$, $Cs_2IrAgCl_6$ and $Cs_2IrAgBr_6$, are direct band-gap semi-conductors, the valence band maximum (VBM) and conduction band minimum (CBM) situated at the X point (0.50, 0.00, 0.50). The $Cs_2RhAgBr_6$, $Cs_2RhAgI_6$ and $Cs_2IrAgI_6$ have indirect band gaps. For $Cs_2RhAgBr_6$, the VBM and CBM situated at the L and X point. However, the VBM and CBM of $Cs_2RhAgI_6$ and $Cs_2IrAgI_6$ are situated at Γ and X point. To gain a better understanding of the origin of the conduction band and valence band, the density TDOS/PDOS of $Cs_2B'AgX_6$ (B' = Rh, Ir, X = Cl, Br and I) are plotted. The valence band is determined by B'-*d* and X-*p* anti-bonding orbitals, while the conduction band is composed of B'-*d* anti-bonding orbitals. The contribution of Cs-*s* anti-bonding orbitals to the conduction band and valence band is very small. In addition, the main

occupation orbitals of VBM are different for the direct and indirect band gap. For the direct band gap, VBM is mainly occupied by B'-*d* anti-bonding orbital, the X-*p* anti-bonding orbitals are the main occupied states for indirect band gap (B' = Rh, Ir, X = Cl, Br and I).

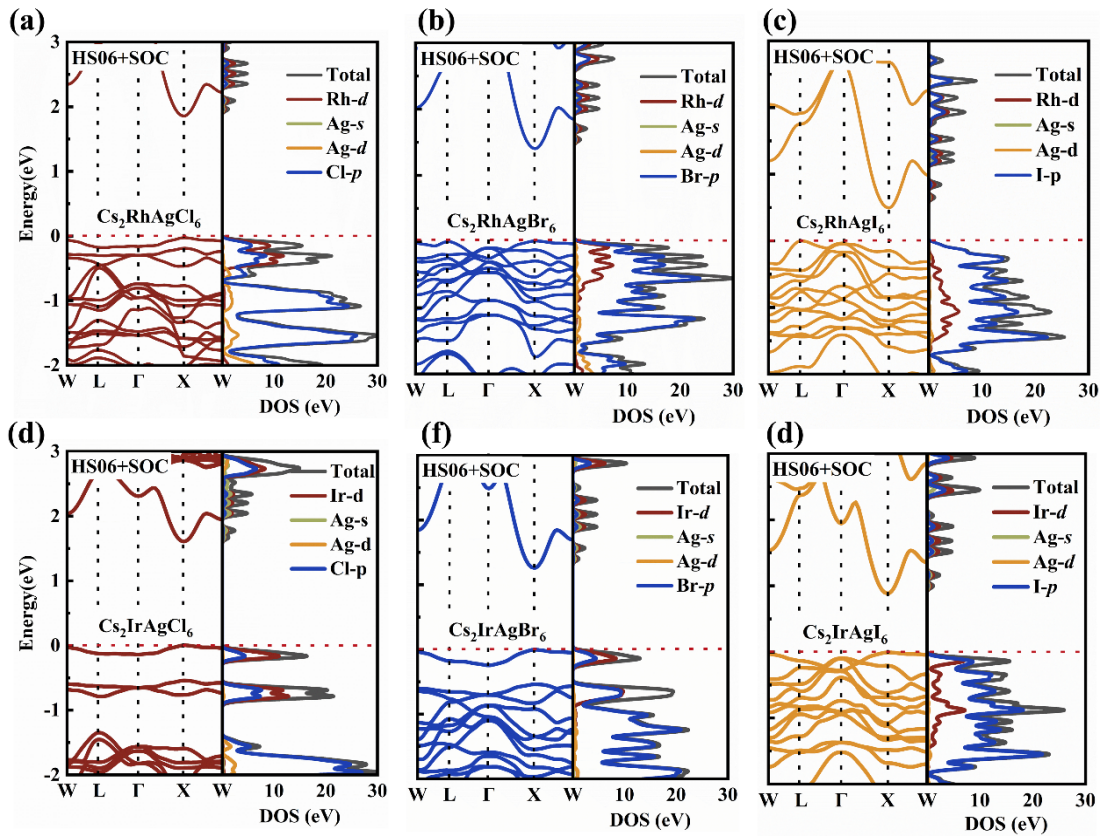

**Fig. 5** HSE06+SOC calculated band structures and DOS for the $Cs_2B'AgX_6$ (B' = Rh, Ir, X = Cl, Br and I) perovskites: (a) $Cs_2RhAgCl_6$, (b) $Cs_2RhAgBr_6$, (c) $Cs_2RhAgI_6$, (d) $Cs_2IrAgCl_6$, (e) $Cs_2IrAgBr_6$ and (f) $Cs_2IrAgI_6$. The fermi-level is set at zero.

The spin-orbit coupling (SOC) effect is strong for elements with larger atomic numbers of Rh, Ir and Ag. Therefore, SOC should be considered to further improve the accuracy. In **Fig. 5**, the band structures and DOS are also calculated using HSE06+SOC functionals. The band gaps of perovskites $Cs_2B'AgX_6$ (B' = Rh, Ir, X = Cl, Br and I) calculated with HSE06+SOC functional is smaller than that with HSE06, it is mainly attributed to the SOC which induced splitting of the Rh-4*d* and Ir-5*d* valence bands. The valence band splitting of $Cs_2IrAgX_6$ is more obvious than that of $Cs_2RhAgX_6$. This can also be observed from the band gap change. Considering the SOC effect, the band gap ranges of $Cs_2RhAgX_6$ and $Cs_2IrAgX_6$, are changed to 0.24-0.5 eV and 0.24-0.12 eV after using the HSE06+SOC functional.

## 3.3 Optical properties

The light absorption coefficient is the most important parameter for assessing the optical properties of material. The high light absorption coefficient facilitates the excitation of electrons from the valence band (VB) to the conduction band (CB). However, for PBE functions, the optical band gap is usually underestimated and the optical absorption coefficient overestimated. Therefore, we investigated the theoretical optical properties of $Cs_2B'AgX_6$ (B' = Rh, Ir, X = Cl, Br and I) by using HSE06 functionals. The absorption coefficient can be obtained from dielectric functions, which are usually used to describe the linear response of the crystal system to electromagnetic radiation. The imaginary part $\varepsilon_2(\omega)$ of the dielectric function is derived from the appropriate momentum matrix elements between the occupied and the unoccupied wave functions within the selection rules over the Brillouin zone, and the real part $\varepsilon_1(\omega)$ of dielectric function follows the Kramer-Kronig relationship [47-51]. Thus, the absorption coefficient $\alpha(\omega)$ is determined by the following equation [52]:

$$\alpha(\omega) = \sqrt{2}\frac{\omega}{c}\sqrt{\sqrt{\varepsilon_1(\omega)^2 + \varepsilon_2(\omega)^2} - \varepsilon_1(\omega)} \tag{5}$$

The absorption coefficient $\alpha(\omega)$ is the outcome of the interaction between photons and valence band electrons, which determines the maximum infiltrated depth of photons with a particular energy before being absorbed. The comparative absorption spectra of $Cs_2RhAgX_6$ and $Cs_2IrAgX_6$ (X = Cl, Br and I) are shown in **Fig 6 a** and **Fig 6 b**. We found that the absorption edge moves towards lower energies as the element at the X-position transitions from Cl to I, which is consistent with the calculated trend of decreasing band gap. The absorption coefficient increases rapidly near the absorption edge to $10^5$ cm$^{-1}$. In **Fig. 6a** and **Fig. 6b**, for the $Cs_2RhAgBr_6$, $Cs_2RhAgI_6$ and $Cs_2RhAgI_6$ systems, they exhibit high optical absorption coefficients in the visible range (1.77-3.18 eV) of $5.1\times10^5$, $5.2\times10^5$ and $2.8\times10^5$ cm$^{-1}$ respectively. Therefore, the $Cs_2RhAgBr_6$, $Cs_2RhAgI_6$ and $Cs_2RhAgI_6$ are suitable to be applied in photovoltaic devices. However, for the $Cs_2RhAgCl_6$, $Cs_2IrAgBrCl_6$ and $Cs_2IrAgBr_6$ systems, the high optical absorption coefficient peaks are located in the UV region, indicating that these systems are suitable for use in UV sensors. The high light absorption coefficients exhibited by $Cs_2RhAgX_6$ and $Cs_2IrAgX_6$ (X = Cl, Br and I) are attributed to the stronger Rh-X, Ir-X and Ag-X bond interactions.

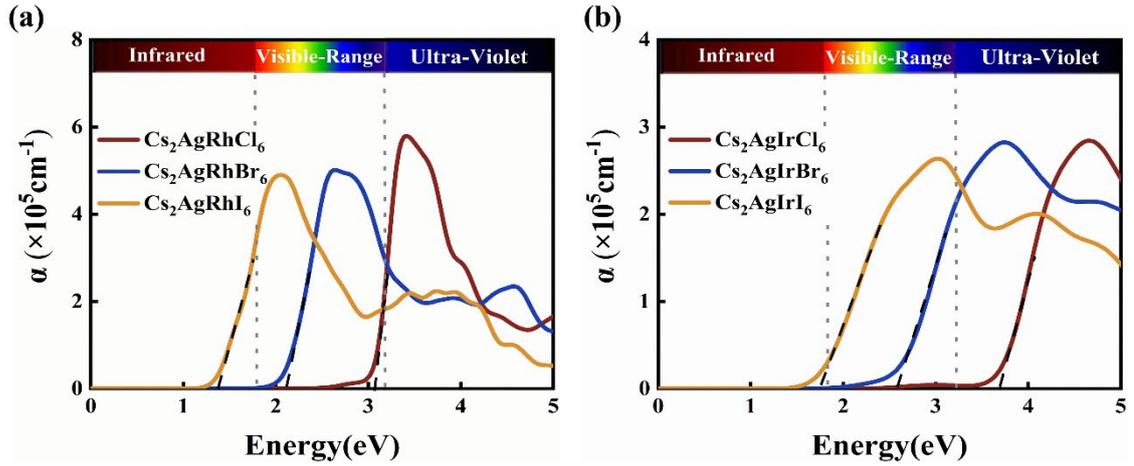

**Fig. 6** The absorption coefficient of (a) Cs$_2$RhAgX$_6$ (X = Cl, Br and I) and (b) Cs$_2$IrAgX$_6$ (X = Cl, Br and I) by HSE06 functional.

## 4．CONCLUSION

In summary, we have investigated the structural, electronic and optical properties of Cs$_2$B'AgX$_6$ (B' = Rh and Ir, X = Cl, Br and I) perovskites using first-principles calculations. Our results show that Cs$_2$B'AgX$_6$ (B' = Rh and Ir, X = Cl, Br and I) perovskites have a larger decomposition energy, showing an enhanced stability. By substitution of halogen elements (from I to Cl), geometry, thermal and mechanical stability tend to be more stable. The Cs$_2$RhAgCl$_6$, Cs$_2$RhAgBr$_6$, Cs$_2$IrAgCl$_6$ and Cs$_2$IrAgBr$_6$ have direct band gaps in the range from 0.9-1.8 eV. In addition, SOC has strong effect in Cs$_2$IrAgX$_6$ and reduced more band gap than Cs$_2$RhAgX$_6$ (X = Cl, Br and I). These double perovskites also a strong optical absorption and promising applications in the optoelectronic devices.

**Notes**

There are no conflicts to declare.

**ACKNOWLEDGMENTS**

This work was supported by the Natural Science Basic Research Program of Shaanxi (No. 2021JM-371), Fund of State Key Laboratory of IPOC(BUPT) (No. IPOC2019A013), and Open-Foundation of Key Laboratory of Laser Device Technology, China North Industries Group Corporation Limited (Grant No. KLLDT202103).